\documentclass[conference]{IEEEtran}
\IEEEoverridecommandlockouts
\usepackage{cite}
\usepackage{amsmath,amssymb,amsfonts}
\usepackage{algorithmic}
\usepackage{graphicx}
\usepackage{textcomp}
\usepackage{xcolor}
\def\BibTeX{{\rm B\kern-.05em{\sc i\kern-.025em b}\kern-.08em
    T\kern-.1667em\lower.7ex\hbox{E}\kern-.125emX}}
\begin{document}

\title{Github Data Exposure and\\ Accessing Blocked Data using\\ GraphQL Security Design Flaw} 

\author{\IEEEauthorblockN{Shahriar Yazdipour}
\IEEEauthorblockA{\textit{} \\
\textit{Technische Universität Ilmenau}\\
Ilmenau, Germany \\
shahriar.yazdipour@tu-ilmenau.de}
}

\maketitle

\begin{abstract}
This research study was conducted to illustrate how it is easily possible to get data access to disabled or blocked repositories in Github using GraphQL. There are situations in which you can lose access to your Github repositories; When you use the paid version of Github services and do not pay the monthly payment or another situation is that when you use Github from the countries in the United States sanction list. Having an insecure repository with malicious usages can also put your repository in Github blacklist. In all of these situations, Github will block and disable your repository and you will lose access to your files, codes and project assets. Here, we will discuss the procedure of how an Ethical Hacker can gain access to all those blocked data with GraphQL functionality.
\end{abstract}


\section{Introduction}

On July 2019, GitHub, the world’s largest host of source code, confirmed that they blocked private repositories of users in Iran, Syria, and Crimea after blocking developers from Cuba and North Korea, due to the impact of the U.S. trade restrictions and sanctions. After this, developers from these countries lost their access over their codes and files on Github Servers. Similarly, organizations that do not pay their monthly payment, Github will disable their repositories too. 

In this security research paper, we will illustrate how did we investigate and collect data from Github Mobile Application to find GitHub private GraphQL queries and how simply users can bypass the Github security check to gain their inaccessible data back. Also, we will discuss ways of overcoming these issues.

\section{APIs and API Abuse}

A contract between a client and a server is called an API. The most popular API abuse is begun with the fact that each side fails to honor their side of the contract. In this scenario, the client violates the server by assuming its behavior\cite{chess2007secure}. It is also possible to violate the client-server contract if the client could access unwanted information on the server-side.

Two of the main types of well-known APIs are REST API\cite{masse2011rest} and GraphQL. Recently GraphQL is introduced and open-sourced by Facebook. It is a conceptual framework for giving a new model of data access interface on the Web which is increasingly embraced in online services\cite{hartig2017initial}. The framework involves an original graph query language and a query compiler on each side of an API service (caller and callee) to request and understand the required data. 

Github is starting to adopt GraphQL as an alternative way to access their web services. Github GraphQL has adequate documentation, although Github implementation of GraphQL does not seems to have all of the features of their RESTful API right now.


\section{Data Gathering}

The initial step toward finding a way to gain access is collecting data. The new Github iOS Mobile Application which is now in Public Preview is the only possible way to see disabled repositories directory. The app can not open Binary files and download any file. Thus, by capturing the mobile application communication data we can understand the process of accessing repository structure and non-binary files, but not binary files. After analyzing transferred data we will try to replicate those calls on our application to make backup files from lost data.

Using a well-known penetration test\cite{arkin2005software} method called "Man In The Middle" let us set up a proxy between the application and Github Server to collect data transfer between them. Pen-testing is necessary for recognizing non-functional access controls.
A security attack that the hacker secretly transfers and probably modifies the communication between two parties who assume they are directly talking with each other,  is called the Man-In-The-Middle or MITM attack \cite{callegati2009man}.

Initially, we start with popular client applications, Github Desktop and FastHub for Android, but all of them are based on Github REST API. In our knowledge, Github RESTful API is well tested and it is not possible to gain unwanted data and requests end up failing after trying to access disabled repositories.

However, after publishing the first public preview version of the Github iOS application, We immediately find the difference which is the possibility to see repositories directory and GraphQL backend to communicate with the server in this application.
To set up the MITM environment for our scenario, we add a proxy in the middle of Mobile client and Github servers using a reputable penetration tool called Burp-Suite\cite{rahalkar2016hacking} and due to the use of SSL for network communication inside the application, we installed CA Certificate (Certification Authority) via iOS Trusted Profile. After successfully capturing communication packets, we collect data related to disabled repositories and we discovered there is a lack of security check and policy control in GraphQL.


\section{Data Understanding}

GraphQL is a newly introduced a conceptual framework for implementing a new way of data access interface with a new graph query language \cite{hartig2018semantics}. By analyzing the requests and responses data, we can breakdown the messages into two main parts, header and body.

Inside the header, we can find request method, host and destination, GraphQL Apollo Operation or Function name, Authorization and other information.

\begin{verbatim}
POST /graphql HTTP/1.1
Host: api.github.com
Content-Type: application/json
X-APOLLO-OPERATION-NAME: RepoFiles
GraphQL-Features: pe_mobile
User-Agent: GitHub/1.0.0
Authorization: Bearer XXXX
Accept-Encoding: gzip, deflate
Accept-Language: en-US;q=1.0
\end{verbatim}

And in the body segment, we can see the GraphQL the request or the response of the request in JSON. Inside request body \verb|operationName| clarifies which GraphQL operation must run, \verb|query| is our demanded GraphQL query to provide data and \verb|variables| declares variables which it is required in our query. And inside the response body, we receive data which requested data asked via GraphQL query.

\begin{figure}[hbt!]
  \includegraphics[width=\linewidth]{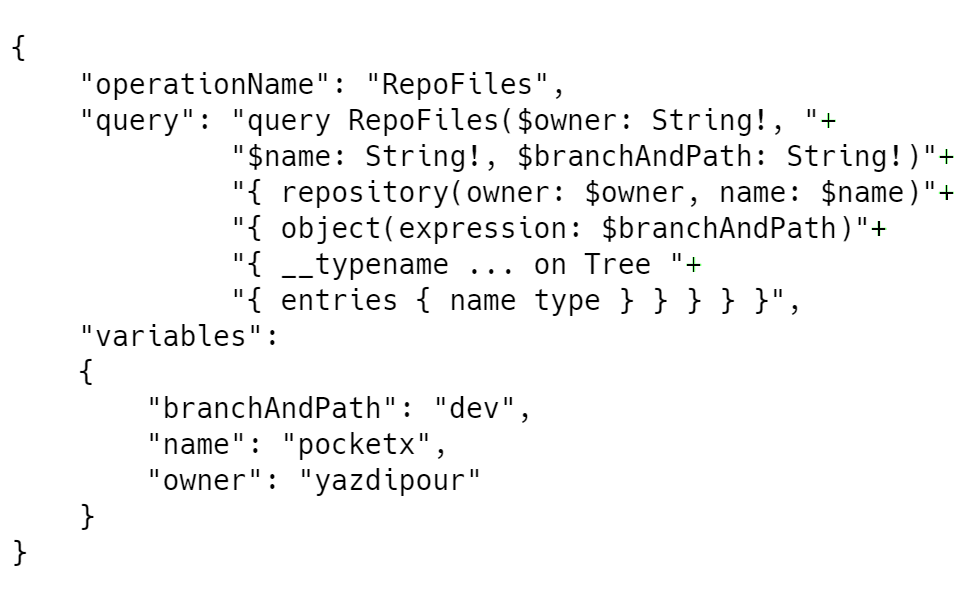}
  \caption{JSON Request containing GraphQL Query}
  \label{fig:carbon}
\end{figure}

\section{Data Access}

\begin{figure}[hbt!]
  \includegraphics[width=\linewidth]{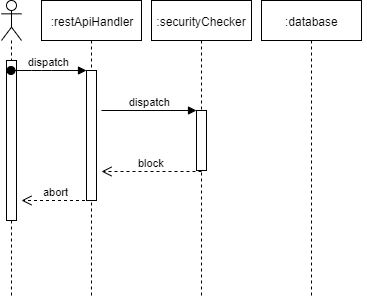}
  \caption{RestApi Sequence Model}
  \label{fig:diagram_rest}
  \includegraphics[width=\linewidth]{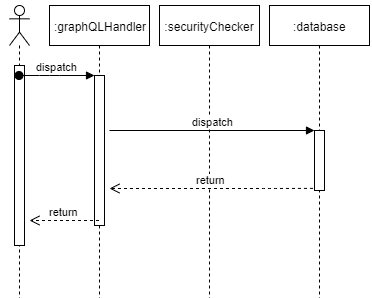}
  \caption{GraphQL Sequence Model}
  \label{fig:diagram_ql}
\end{figure}

As our result shows, when trying to access the disabled repository, REST API abort our request. In our assumption, there must be a layer of security and policy check between database and REST API service, which handles security tasks.[Figure\ref{fig:diagram_rest}]. 

According to our analysis, using GraphQL queries will not be rejected by GitHub servers when trying to reach disabled repositories. Our research shows that GraphQL service is not using the same security layer as REST service uses. So this will allow us to bypass security checks and access blocked files and disabled repositories [Figure\ref{fig:diagram_ql}].

Thus with the aid of Official Github GraphQL documentation, we can develop a program to access the directory of disabled repositories. 

But what about files? Github divides Files or Blobs into two distinct types, Non-Binary and Binary files. Non-binary blobs are records like texts and codes, but pictures, exe, dll, ... files are considered binary files. Now we will discuss how it is possible to access these two types of files.

\subsection{Non-Binary Files}\label{AA}

Using Public Github GraphQL features, it is possible to access non-binary data easily. We are only required to cast the object to blob and call its' \verb|text| property. This query will return the encoded blob content inside the response JSON. After decoding the response we can save it in memory.

\begin{figure}[hbt!]
  \includegraphics[width=\linewidth]{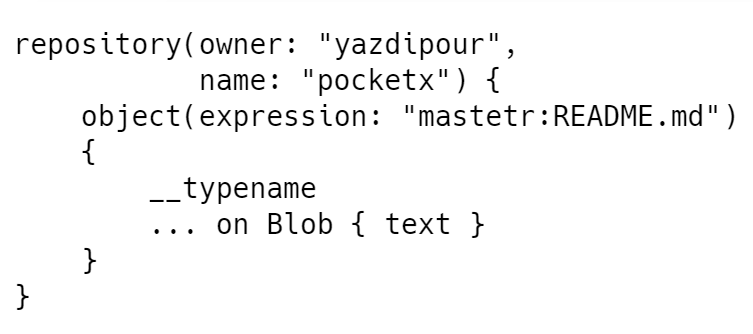}
  \caption{GraphQL Query to access Non-Binary Files}
  \label{fig:carbon2}
\end{figure}

\subsection{Binary Files}

After our initial attempt to access binary files, we discovered due to unknown reasons the mobile application is reaching the files using their URLs directly, which we can not access. For more investigation, we downgraded our mobile application version to the very first public preview build. We observed that at that version, GitHub tried to get binary files using \verb|"file(path){contentHTML}"| GraphQL query which is not mentioned inside Github Online Documents. We decided to use this function with our access token, taken from Github Developer settings, but it did not work. It appears, Github is using this private function for its application and it is not available for the public.

For each server request via Github API, Access Token for Authentication is needed. Inside captured transferred data, in the header section of the request is a line for Authorization. Which this section ends with  \verb|"Bearer"| and a forty characters length Access Token or Client Secret that this is the token Github generates for developers or applications gets it after users authorize themselves.

Assume that authentication token is available, Here, we adopted OctoKit.GraphQL.Net Library\footnote{OctoKit.GraphQL.Net is the Official client library for Github GraphQL for DotNet.}, to write this piece of code for generating GraphQL Queries and sending query requests to Github servers. To apply the unreleased and unsupported private function of Github GraphQL, it is easily possible to force our function into query by simply injecting string inside the compiled query.

\begin{figure}[hbt!]
  \includegraphics[width=\linewidth]{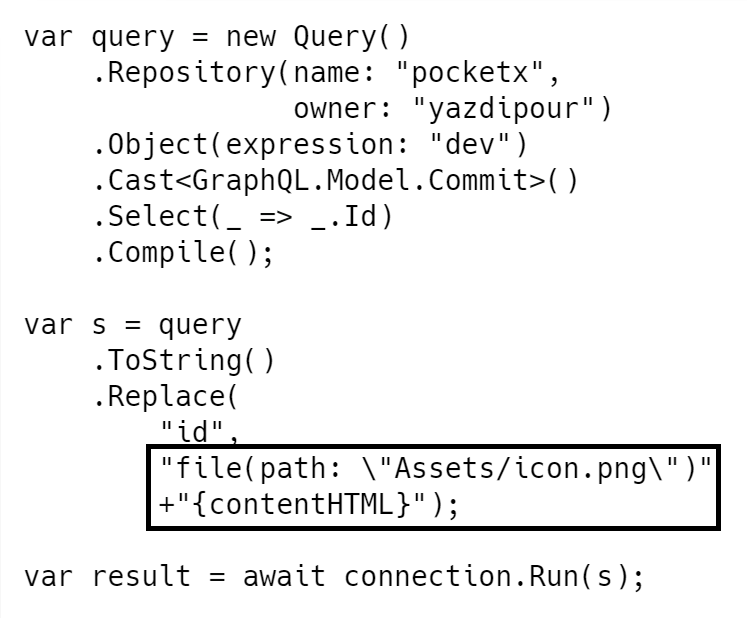}
  \caption{C\# Code Accessing Image File}
  \label{fig:carbon3}
\end{figure}

Quickly, with Breadth-first traversing in the directory tree and downloading files, we can fully restore our disabled repository.

\subsection{Authentication Token}

In our security research, we found out that there are two types of Client Secret Tokens. One which is publicly available for developers and applications to use, and the private tokens which Github is using for it their Mobile Application to access GraphQL private functions.
The \verb|"file"| function is a private GraphQL function that allows Github Mobile Application to get binary files inside the server response. Using this query in our system is only possible by high jacking Github Access Token.

Accessing the token is possible by mimicking the exact behavior of the Github Official Application. If we track the login process in the mobile application we can figure out that \verb|/login/oauth/access_token| route will give us Client Secret after user login to their account if we provide proper \verb|"client_id"| and \verb|"code"|. Another important thing that we have to take into consideration is that GitHub servers will return authentication data to users' mobile to a specific URI or Activation Protocol, in which the Operating System will respond to the specific application on the phone which is registered with that protocol. In this case, Github shows the redirecting URI on the login page, and in this case, it is \verb|github://com.github.ios|. So, by having all this information, we can develop a twin application, with the authentication process, to get the Security Token.
\section{Patch}

In the following years, security flaws similar to this problem probably will be seen more than often and especially with the growth in usage of GraphQL among developers and increase in GraphQL adoption in API services. 
    We have to be alert while transporting security and policy checks from RESTful API to GraphQL. 

One of the fundamental solutions for this specific issue could be, Handling security checks on the Database Management System side. Depending on the structure and architecture of the system, this might be efficient for some systems due to the separation of sub-systems or it might not be proper for specific cases. 

Alternative method would be adding another layer after REST API and GraphQL layers, therefore a central subsystem layer handles data transition so that it will be possible to dictate our policies directly on this layer.

Encryption of information at rest and in transition can benefit you comply with information security regulations\cite{owasp2017top}. "End-to-end" encryption method can be the ultimate solution for such matters. Although, Limitations of this method have been already known; The End-to-end encryption approach is not enough to stop compromising of data in a network that operates on untrusted sub-net processors\cite{padlipsky1978limitations}. Also, timing and message lengths of deliveries will have a straight effect on the speed of receiving and decryption of data. Therefore end-to-end encryption is beneficial for Super Confidential scenarios, however, it may not be practically efficient for general purposes.

OWASP or The Open Web Application Security Project\cite{owasp2017top} is a non-profit organization committed to providing honest, realistic information about application safety and security. The "OWASP Top 10 Web Application Security Risks"\cite{owasp2017top} guides programmers and security experts on the common important vulnerabilities which are generally observed in web applications, which are also simple to exploit.

The fifth most commonly security issue inside OWASP Top 10 list is "Broken Access Control". It means, Inadequately configured constraints on verified users enable them to obtain unauthorized information or functionality, such as inspecting sensitive documents and access rights. Hackers can exploit and get unauthorized access and/or data, such as access to other users' records, view sensitive files, etc \cite{owasp2017top}. Now we can assume that this Github GraphQL security problem is in the category of Broken Access Control.  

\section*{Conclusion}

In conclusion, with studying the communication traffic between the application and Github servers, we were able to access Github private GraphQL queries and restore all of the files inside our disabled repository as a result of a security design flaw in Github GraphQL service.

It is now clear that soon, many security problems will be witnessed similarly in other services. Security researchers must be concerned about how to find and prevent such problems from happening in the future. It is hoped that this research will excite further investigations in this area.


\section*{Acknowledgment}

We would like to gratefully acknowledge the Github Security Team for all their support and professionally handling this issue after reporting it.

\newpage
\bibliographystyle{unsrt}
\bibliography{ref}

\end{document}